\documentclass[10pt]{article}

\usepackage{amssymb}
\usepackage{graphicx}
\usepackage[numbers,compress]{natbib}
\begin{document}

	\title{New one-parameter models of dynamical particles in spatially flat FLRW space-times}
	
	\author{Ion I. Cot\u aescu\thanks{Corresponding author E-mail:~~i.cotaescu@e-uvt.ro}\\
		{\it West University of Timi\c soara,} \\{\it V. Parvan Ave. 4,
			RO-300223 Timi\c soara}}
	
	\maketitle

\begin{abstract}
New one-parameter models of non-rotating dynamical particles are derived as isotropic solutions of  Einstein's equations with perfect fluid in space-times with FLRW asymptotic behaviour  generalizing thus the  models  proposed recently  in [I. I. Cot\u aescu, Eur. Phys. J. C  (2022) 82:86].   These particles are produced by central singularities of the fluid density but without changing the pressure of the asymptotic FLRW space-times. The principal features of these models are investigated  using a brief graphical analysis for pointing out the role of the new free parameter. The conclusion is that this gives rise to families of models which behave as non-rotating black holes in the physical space domain bordered by the black hole and cosmological horizons.

Pacs: 04.70.Bw
\end{abstract}


\section{Introduction}

 The Friedmann-Lema\^ itre-Robertson-Walker (FLRW) space-times play an important role in  actual cosmology as  plausible models of our universe in various epochs of evolution. These space-times evolve according to their scale factors  which solve the Einstein-Friedmann equations with the energy-momentum tensor of an isotropic perfect fluid. For building more realistic models one populates the FLRW manifolds either with static black holes, which are vacuum solutions of the Einstein equations \cite{BH}, or looking for dynamical particles defined as exact solutions of Einstein's equations with perfect fluid laying out a FLRW asymptotic behaviour (see for instance Ref. \cite{TB1}).

A prominent model of non-rotating dynamical particles was proposed by McVittie \cite{MV} and then studied by many authors  mainly in physical frames with Painlev\' e-Gullstrand coordinates \cite{Pan,Gul} where their properties are visible and intuitive \cite{MV1,MV2,MV3}. Recently the McVittie geometry was generalized to new solutions of the Einstein equations with perfect fluid behaving asymptotically as FLRW space-times with curved space sections \cite{MV4}. The general feature of these models is that their gravitational source is the fluid pressure which is singular on the Schwarzschild sphere while the density remains that of the asymptotic FLRW space-time.

As in our opinion the McVittie dynamical particles  cannot be seen as a natural generalizations of the Schwarzschild black holes which are produced by central singularities, we proposed recently  a new type of dynamical particles which are exact solutions of Einstein's equations with perfect fluid preserving the fluid pressure of the asymptotic FLRW space-times  but giving rise to  central point singularities in their densities \cite{Cot}. We have shown that these dynamical particles behave as non-rotating black holes in the physical space domain bordered by the black hole and cosmological dynamical horizons. Moreover, these dynamical particles give rise to photon spheres and black hole shadows as in the case of the static Schwarzschild black holes.  For this reason we say here that these are  Schwarzschild-type dynamical particles or simply point dynamical particles.

We obtained these new solutions without using the traditional Schwarzschild binomial $1-\frac{2m}{r}$  from which we kept only the last term that helped us to construct the metrics of the dynamical particles of Ref. \cite{Cot} in a familiar manner. In this paper we would like to continue this study showing how these models  can be generalized without compromising their FLRW asymptotic behaviour. We obtain thus more complicated models that depend on a new free parameter $\kappa$ determining their principal features including the evolution of their black hole and cosmological horizons. We would like to briefly study these models we refer here  simply  as $\kappa$-models.

We start in the next section  presenting our method of deriving solutions of Einstein's equations with perfect fluid in physical proper frames. The next section is devoted to the McVittie and Schwarzschild-type dynamical particles whose properties are compared for understanding the differences between these two types of models. Our new solutions are presented in Section  4 where we show that the $\kappa$-models can be constructed either giving the mass function and the parameter $\kappa$ or postulating the asymptotic behavior specifying the constant $\kappa$ and the initial condition for the mass function.   We show that there is a critical instant when a pair of dynamical horizons,  the black hole and cosmological ones,  are arising simultaneously on the same sphere. When the time is increasing these evolve as fold-type curves (or C-curves), the  cosmological horizon, approaching asymptotically to the apparent horizon of the asymptotic FLRW space-time while the black hole one is shrinking to the dynamical Schwarzschild sphere collapsing then to zero. The $\kappa$-models are so complicated that we cannot apply some traditional analytical methods as,  for example,  that of the areal radius. Consequently, we have to rely mainly on numerical and graphical methods as we proceed in Section 5 for investigating the role of the new parameter $\kappa$.  Some concluding remarks are presented in the last section. The method of solving cubic equations is given  in Appendix.

We use the Planck units with $\hbar=c=G=1$.

\section{Dynamical particles in physical frames}

The static space-times with spherical symmetry are studied traditionally in static  frames, $\{t_s, {\bf x}\}$, whose coordinates, $x^{\mu}$ ($\alpha,\mu,\nu,...=0,1,2,3$), are the static time $x^0_s=t_s$ and physical Cartesian space coordinates  ${\bf x}=(x^1,x^2,x^3)$ associated to the spherical ones $(r,\theta,\phi)$. In these frames the line elements have the general  form  
\begin{eqnarray}
ds^2=f(r)\, dt_s^2-\frac{dr^2}{f(r)}-r^2 d\Omega^2 \,,\label{s1s}
\end{eqnarray} 
where $d\Omega^2=d\theta^2+\sin^2\theta\, d\phi^2$. However, for studying dynamical geometries it is convenient to consider the {\em cosmic} or proper time 
\begin{equation}
x^0=t=t_s-\int dr \frac{\sqrt{1-f(r)}}{f(r)}\,,
\end{equation}    
defining the {\em physical} frames $\{t, {\bf x}\}$  of   Painlev\' e-Gullstrand  coordinates  \cite{Pan,Gul}, in which the static line element (\ref{s1s}) becomes
\begin{eqnarray}\label{s}
	ds^2=f(r)dt^2+2\sqrt{1-f(r)}\,dtdr-dr^2 -r^2d\Omega^2\,,\label{ss}
\end{eqnarray}
laying out flat space sections.

On the other hand, for the FLRW space-times ${\frak M}(a)$ of scale factors $a(t)$ one uses the cosmic time and co-moving space coordinates giving the FLRW line element
\begin{equation}
	ds^2=dt^2-a(t)^2 d {\bf x}_c\cdot d{\bf x}_c\,,
\end{equation} 
which in the flat case of $a(t)=1$ becomes  just that of the Minkowski space-time,  ${\frak M}(1)$. The space coordinates of the physical frame  $\{t, {\bf x}\}$ are defined as ${\bf x}=a(t){\bf x}_c$ such that the line element  
\begin{equation}\label{s2}
	ds^2=\left(1-\frac{\dot a^2}{a^2}\, r^2\right)dt^2+2\frac{\dot a}{a}\, r\, dr\, dt -dr^2-r^2d\Omega^2\,,
\end{equation}
depends only on the Hubble function $\frac{\dot a(t)}{a(t)}$  which gives the radius 
\begin{equation}\label{hor}
	r_a(t)=\left|\frac{a(t)}{\dot{a}(t)}\right|\,,
\end{equation}
of the dynamical apparent horizon.   

For example, the Schwarzschild-de Sitter (SdS) black hole is defined by  the function
\begin{equation}\label{frSdS}
f_{SdS}(r)=1-\frac{2m}{r}-\omega_{dS}^2 r^2\,,\quad \omega_{dS}=\sqrt{\frac{\Lambda}{3}}\,,
\end{equation}  
which depends on the static black hole mass $m$ and  the Hubble-de Sitter constant $\omega_{dS}$ related to the cosmological constant $\Lambda$ determining the asymptotic behaviour of the background.  Indeed, for $r\to\infty$ we have $f_{SdS}\to 1-\omega_{dS}^2 r^2$ such that the metric (\ref{s}) takes the form (\ref {s2}), of the de Sitter expanding universe ${\frak M}(a_{dS})$ with the scale factor $a_{dS}(t)=\exp(\omega_{dS}\,t)$.  

In what follows we focus on space-times of non-rotating dynamical particles  (or black holes) with spherical symmetry whose physical frames have  line elements of the general form
\begin{equation}\label{s1}
ds^2=g_{00}(t,r)dt^2+2g_{0r}(t,r)dr\, dt-g_{rr}(r)dr^2-r^2d\Omega^2\,,
\end{equation}
depending on metric tensors  that solve  the Einstein equations
\begin{equation}\label{Ein}
G^{\mu}_{\,\nu}=\Lambda\delta^{\mu}_{\nu}+ 8\pi\left[ (\rho+p)U^{\mu}U_{\nu}-p\delta^{\mu}_{\nu}\right]\,,
\end{equation}
with a perfect fluid of density $\rho$ (of matter or energy) and pressure $p$, moving with the four-velocity $U^{\mu}$ with respect to the physical frame under consideration. In a proper co-moving frame where the four-velocity has the components 
\begin{equation}
U_{\mu}=\left(\frac{1}{\sqrt{g^{00}}},0,0,0\right)\,, \quad U^{\mu}=g^{\mu0}U_0=\frac{g^{\mu 0}}{\sqrt{g^{00}}}\,,
\end{equation}
Eqs. (\ref{Ein}) are solved by an isotropic Einstein tensor  whose non-vanishing components satisfy
\begin{eqnarray}
&&G^r_r=G^{\theta}_{\theta}=G^{\phi}_{\phi}\equiv G\,,\label{iso}\\
G^0_r=0 &\Rightarrow&G^r_0=\frac{g^{0r}}{g^{00}}\left(G^0_0-G\right)\,,\label{cond}
\end{eqnarray}
defining the gravitational sources  in the co-moving frame as
\begin{eqnarray}
	G^0_{0} &=&\Lambda + 8\pi \,\rho\,, \label{E10}\\
	G&=&\Lambda - 8\pi \, p \,.\label{E20}
\end{eqnarray}
Moreover, the asymptotic manifold must also be a solution of the equation (\ref{Ein}) corresponding to the asymptotic gravitational sources.  In general, this is a spatially flat FLRW  space-times  $(M,a)$  as  for $r\to\infty$  the line element (\ref{s1}) is supposed to take the asymptotic form (\ref{s2}).     The corresponding Einstein tensor has  diagonal elements satisfying the Friedmann equations
\begin{eqnarray}
G^0_{0}(a) &=&3\,\frac{\dot{a}^2}{a^2}=\Lambda + 8\pi \,\rho_a\,, \label{E1}\\
G(a)&=& 2\frac{\ddot{a}}{a} + \frac{\dot{a}^2}{a^2}=\Lambda - 8\pi \, p_a \,,\label{E2}
\end{eqnarray}
and a non-diagonal one,  $G^r_0(a)$, satisfying the condition (\ref{cond}). The gravitational sources of the asymptotic space-time ${\frak M}(a)$   are the asymptotic density and pressure, $\rho_a=\lim_{r\to\infty}\rho$ and respectively $p_a=\lim_{r\to\infty}p$.

In this framework many models of dynamical particles that may behave as black holes are considered so far in various geometries (as presented in Ref. \cite{TB1}).

\section{McVittie and Scwarzschild-type dynamical particles}

Of a special interest is  the McVittie \cite{MV} class of metrics describing isotropic dynamical particles in space-times ${\frak M}(a,m)$ whose line elements in physical frames have the form (\ref{s1}) with  
\begin{eqnarray}
g_{00}(t,r)&=&1-\frac{2m}{r}-\frac{\dot a(t)^2}{a(t)^2}r^2\,,\label{met1}\\
g_{0r}(t,r)&=& \frac{\dot a(t)}{a(t)}\frac{r}{\sqrt{1-\frac{2m}{r}}}\,,\qquad ~~~g_{rr}(r)=\frac{1}{1-\frac{2m}{r}}\,,\label{met2}
\end{eqnarray}
depending only on the static mass $m$ and the Hubble function given by the scale factor $a$ of the asymptotic FLRW space-time  ${\frak M}(a)$ with the line element  (\ref{s2}) which is just the asymptotic limit for $r\to \infty$ of the McVittie one  \cite{MV4}.  In other respects, in the static limit, when $a(t)\to 1$ and $\dot a(t)\to 0$,  the McVittie line element becomes the Schwarzschild  one of static frame, 
 \begin{equation}\label{asymMv}
	ds^2=f_S(r) dt^2-\frac{dr^2}{f_S(r)}-r^2 d\Omega^2\,,\quad f_S(r)=1-\frac{2 m}{r}\,,
\end{equation}
but depending on the cosmic time $t$ instead of the static one $t_c$. In addition we observe that the McVittie metrics in physical frames lay out curved space sections.

In  spaces ${\frak M}(a,m)$ the dynamical particles behave as black holes but only in the physical domains where the metric component (\ref{met1}) is positively defined, $g_{00}(t,r)>0$, and consequently,  the coordinate $t$ is the cosmic time. Solving the cubic equation   $g_{00}(t,r)=0$ as in the Appendix we find that in the general case of arbitrary functions $a(t)$ the real solutions arise only in  a given time domain. In the case of the expanding space-times there exists a critical time $t_{cr}$ such that for $t>t_{cr}$ we obtain two real solutions with physical meaning, $r_b(t)$ and $r_c(t)$, and another nonphysical one, $r_{np}(t)$, which satisfy  
\begin{equation}
r_{np}(t)<0<r_b(t)<r_c(t)<r_a(t)\,, \quad \forall t>t_{cr}\,,	
\end{equation} 
where $r_a(t)$ is the radius (\ref{hor}) of the apparent horizon of  ${\frak M}(a)$ we call now the {asymptotic} horizon. The physical solutions have complicated and less intuitive forms but for expanding geometries with $r_a(t)\gg m$ increasing monotonously  in time we may write the expansions
\begin{eqnarray}
r_b(t)&=& 2 m  \left[1+\frac{4m^2}{r_a(t)^2}+{\cal O}\left(\frac{m^4}{r_a(t)^4}\right)\right]\  \,, \\
r_c(t)&=&r_a(t)\left[1-\frac{m}{r_a(t)}-\frac{3m^2}{2r_a(t)^2}+{\cal O}\left(\frac{m^4}{r_a(t)^4}\right)\right]\,, 	
\end{eqnarray} 
showing that 	$\lim_{t\to \infty} r_b(t)=2m$ and $\lim_{t\to \infty} r_c(t)=r_a(t)$ 
as in the left panel of Fig. 1.  This behaviour convinces us that $r_b(t)$ is the radius   of the black hole horizon while $r_c(t)$ is that of the cosmological one.  In the physical space domain which appears between the spheres of black hole and cosmological horizons  an observer can measure the black hole behaviour of the dynamical particle.

 { \begin{figure}
		\centering
		\includegraphics[scale=0.34]{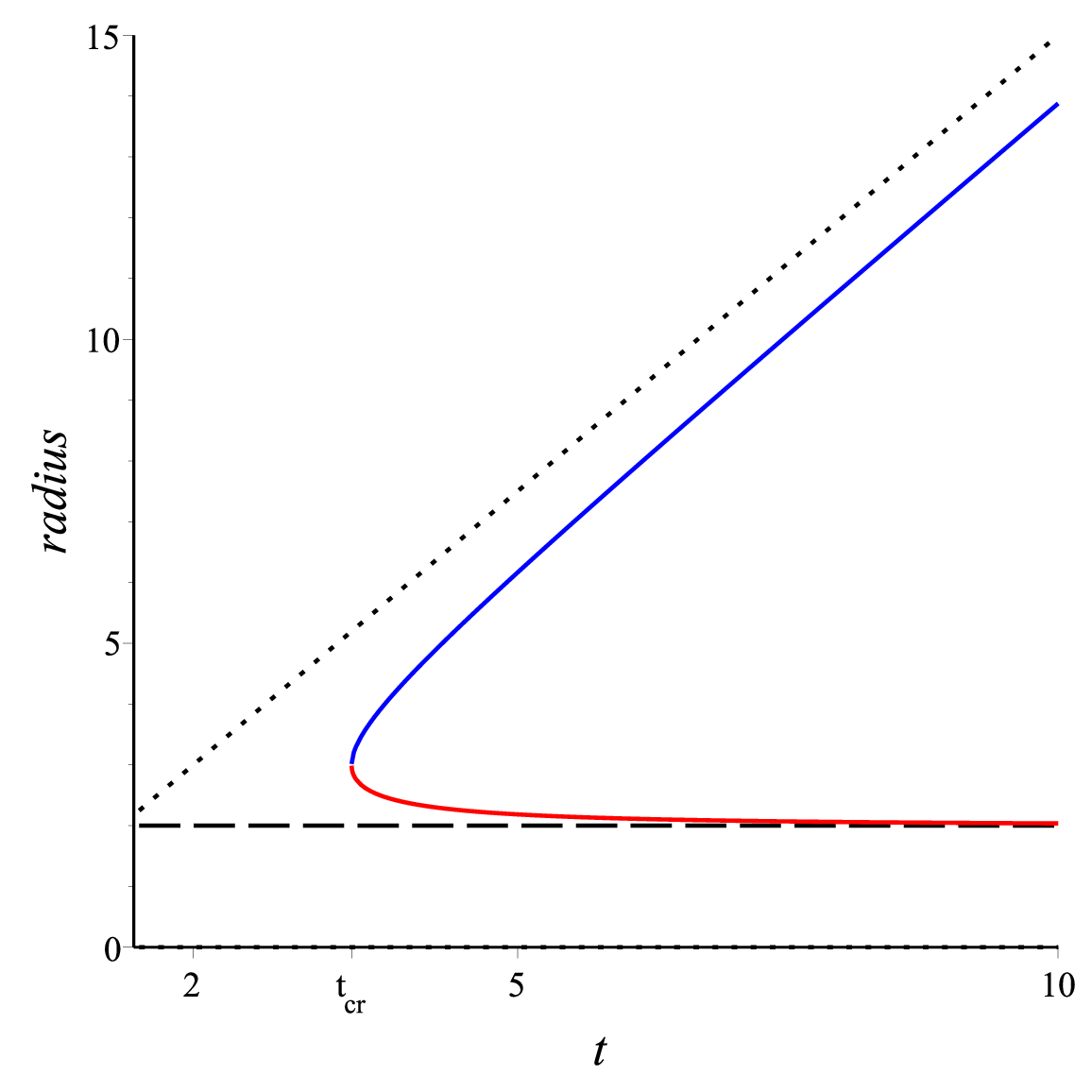}
		\includegraphics[scale=0.34]{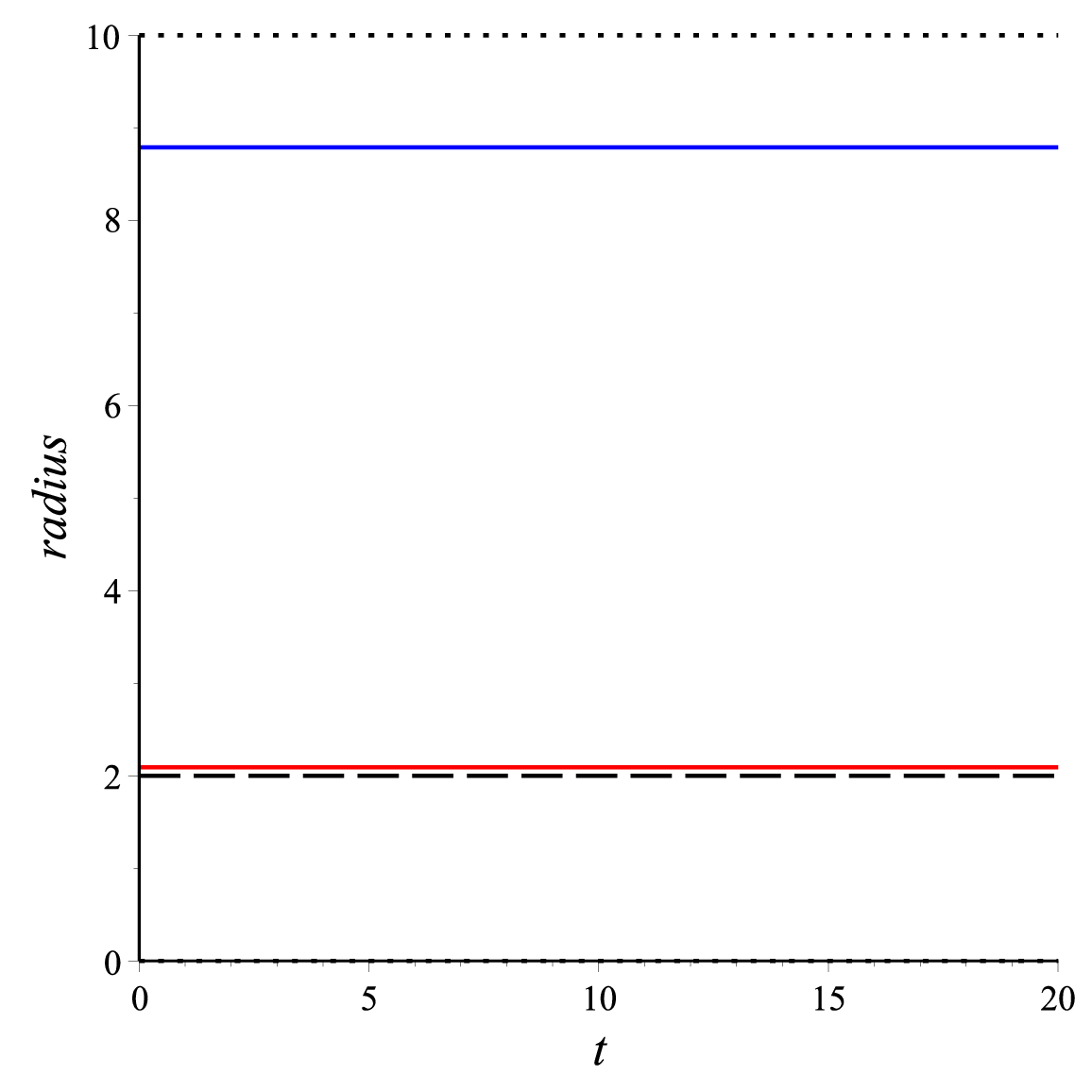}	
		\caption{The evolution of the horizons of McVittie dynamical particles having the asymptotic space-times either the matter-dominated universe (left panel) or the de Sitter expanding universe (right panel). The cosmological horizons are the upper solid lines while the black hole horizons are the solid lower ones. The dotted lines represent the asymptotic horizons and the dashed lines mark the Schwarzschild sphere of radius  $r=2m$.}
\end{figure}}

A notable exception arises when the asymptotic spacetime is the de Sitter expanding universe with $a(t)=\exp(\omega t)$ with the Hubble-de Sitter constant $\omega$. Then  the McVittie metric closes to the Schwarzschild-de Sitter one having the same component (\ref{met1}) with $\frac{\dot a(t)}{a(t)}={\omega}$. In this case  one obtains a static black hole 
with static cosmological and black hole  horizons as in the right panel of Fig. 1.

However, the principal attribute of the McVittie  metric tensor defined by Eqs. (\ref{met1}) and (\ref{met2}) is to be an exact solution of  the Einstein equations  (\ref{Ein})  \cite{MV4} having the diagonal components
\begin{eqnarray}
G^0_0(a,m)&=&G^0_0(a)\,,\\
G(a,m)&=&G^0_0(a)+\frac{1}{\sqrt{1-\frac{2m}{r}}}\left[G(a)-G^0_0(a)\right]\,,
\end{eqnarray} 
while the component $G^r_0$ satisfies the condition (\ref{cond}). Thus the presence of this dynamical particle modifies only the fluid pressure,  which becomes singular on the entire sphere of radius  $r=2m$, but without affecting the density. It is important to observe that this singularity remains outside the physical space domain limited by the black hole horizon of radius $r_b(t)>2m$. These properties suggest that the McVittie dynamical particles  cannot be seen as  genuine Schwarzschild-type ones  that ought to be produced by a central point singularity in a space-time with flat space sections. 

{ \begin{figure}
		\centering
		\includegraphics[scale=0.34]{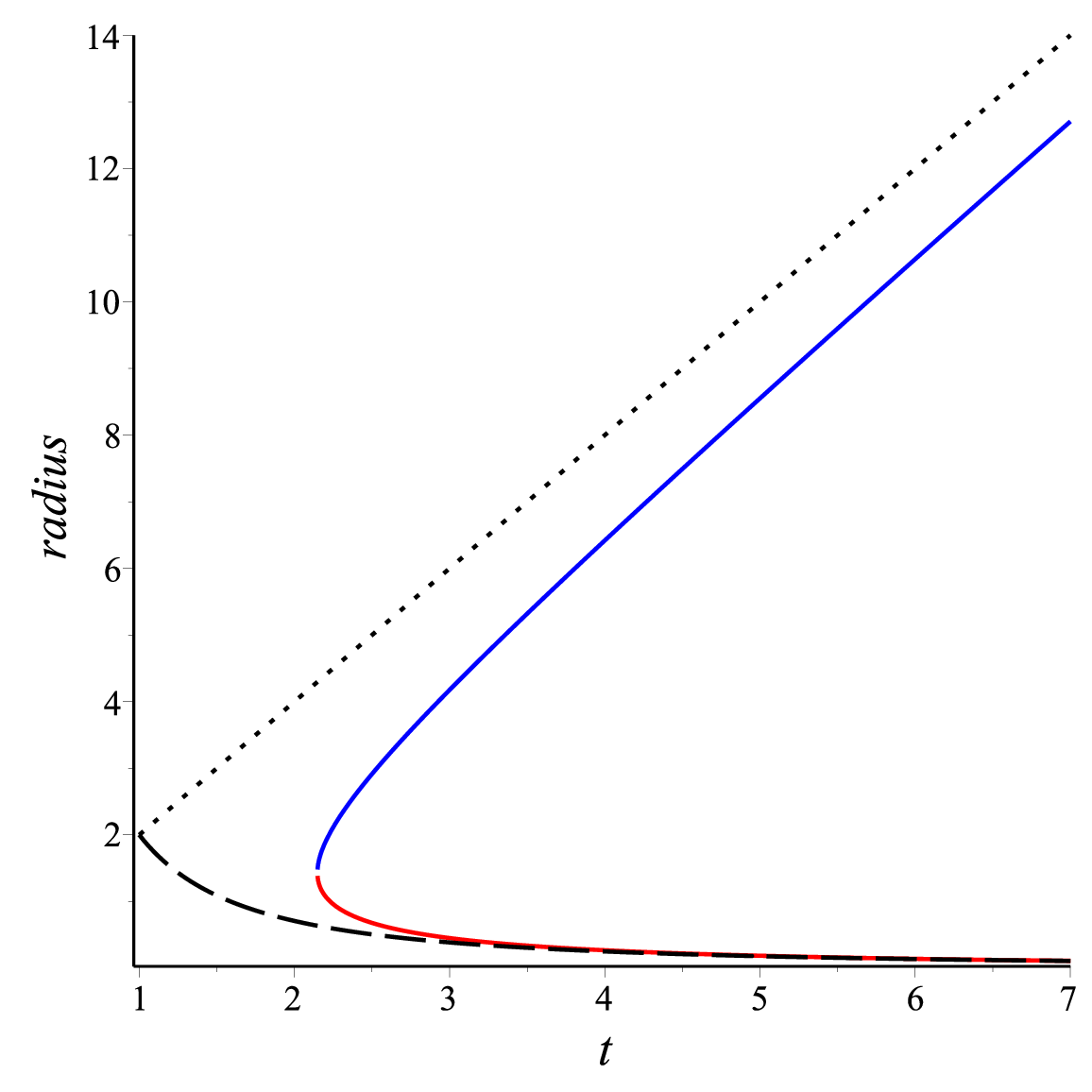}
		\includegraphics[scale=0.34]{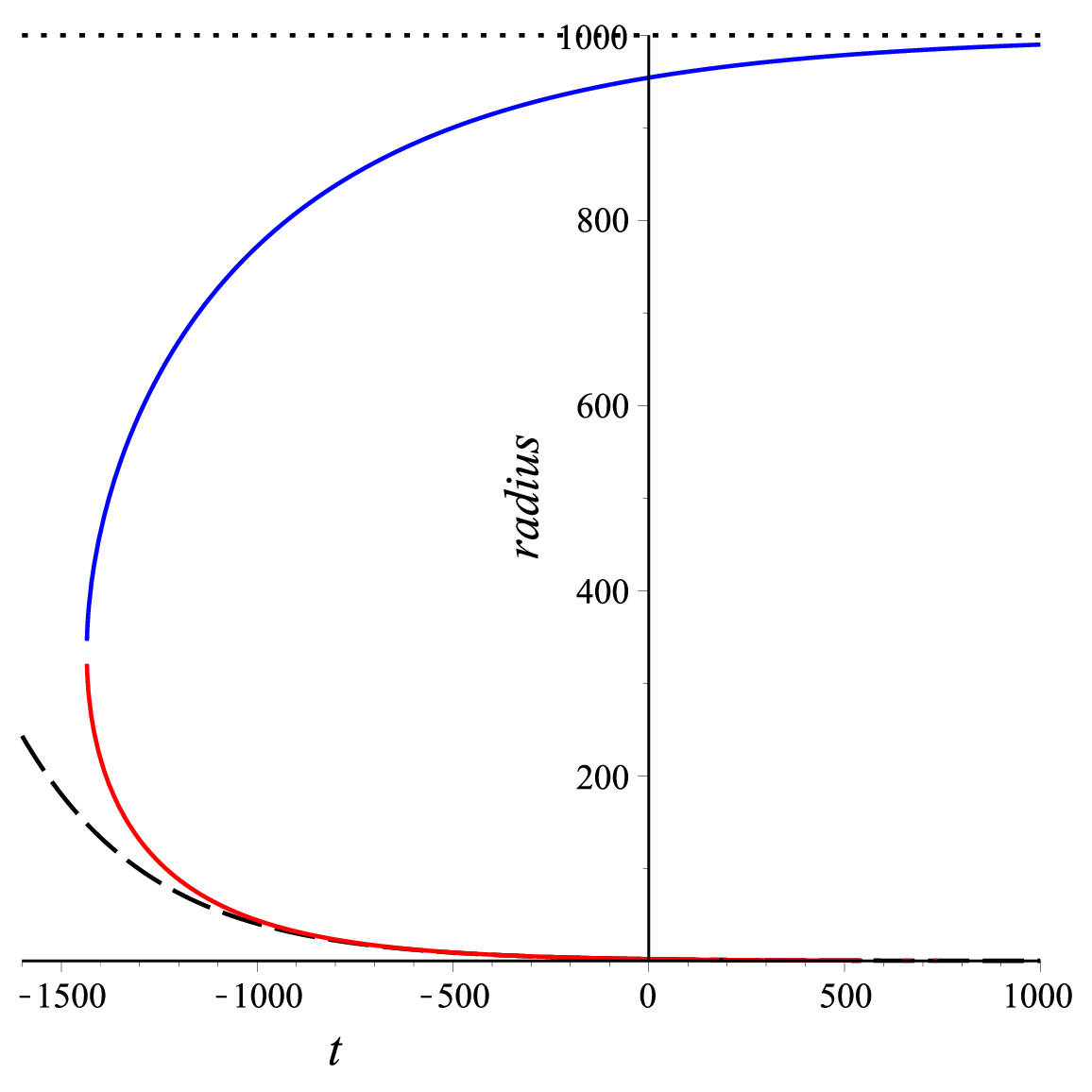}	
		\caption{The evolution of the horizons of Schwarzschild-type dynamical particles having the asymptotic space-times either the matter-dominated universe (left panel) or the de Sitter expanding universe (right panel). The cosmological horizons are the upper solid lines while the black hole horizons are the solid lower ones. The dotted lines represent the asymptotic horizons and the dashed lines mark the radius of the Schwarzschild type dynamical horizon of radius  $r=2M(t)$.}
\end{figure}}

Recently we found a new type of dynamical particles satisfying such exigences. These are   defined  in new space-times with flat space sections, ${\frak M}(a,M_0)$, having physical frames with line elements of the form
\begin{eqnarray}
	ds^2&=&dt^2 -\left[dr-h (t,r)dt\right]^2-r^2 d\Omega^2 \nonumber\\
	&=&\left[1-h (t,r)^2\right]dt^2+2 h(t,r)dr dt -dr^2-r^2 d\Omega^2\,,\label{fam}
\end{eqnarray}
where
\begin{equation}\label{lineBH}
	h(t,r)=\frac{\dot{a}(t)}{a(t)}\,r+\sqrt{\frac{2 M (t)}{r}}\,,\quad  M (t)= \frac{M_0}{a(t)^{
			3}}\,.
\end{equation}
In this metric,  $M(t)$ is the dynamical mass depending on the invariant mass $M_0$ defined as the mass at the initial time $t_0$ when $a(t_0)=1$. This metric is an exact solution of Eqs. (\ref{Ein}) as in the  co-moving physical frames of the space-times ${\frak M}(a,M_0)$ the Einstein tensors have isotropic components,
\begin{eqnarray}
	G^0_{0}(a,M_0) &=&G^0_{0}(a) +\delta\,,\\
	G(a,M_0) &=&G(a) \,,
\end{eqnarray}
where $G^0_0(a)$ and $G(a)$ are the components of Einstein's tensor of the asymptotic manifold ${\frak M}(a,0)\equiv {\frak M}(a)$ as given by Eqs. (\ref{E1}) and (\ref{E2}).  These solutions describe  dynamical particles whose presence gives rise to the additional  terms,
\begin{equation}\label{de}
	\delta(t,r)= 3\frac{\dot a(t)}{a(t)}\sqrt{\frac{2 M(t)}{r^3}}\,,
\end{equation}
which modifies the fluid density  as $8\pi\rho_a ~\Rightarrow~ $ $ 8\pi \rho=8\pi\rho_a+\delta$, introducing a central singularity in $r=0$ while the fluid pressure remains unchanged, $p=p_a$. These properties encourage us to consider these solutions as Schwarzschild-type dynamical particles or point dynamical particles. 

These solutions get  physical meaning only inside the physical space domain delimited by the black hole and cosmological horizons whose radii have to be derived by solving the equation $h(t,r)=1$ for expanding geometries and $h(t,r)=-1$ for collapsing ones \cite{Cot}.   As in the previous case, there is a critical instant $t_{cr}$  such that for  $t>t_{cr}$  we have two positive solutions giving the horizon radii $r_b(t)<r_c(t)$.  For understanding their role we considered  the expansions \cite{Cot}
\begin{eqnarray}
	r_{b}(t)&=&2M(t)\left[1+\frac{4 M(t)}{r_a(t)}+{\cal O}\left(\frac{M(t)^2}{r_a(t)^2}\right) \right]\,,\label{rb}\\
	r_{c}(t)&=&r_a(t)\left[1-\sqrt{\frac{2M(t)}{r_a(t)}}-\frac{M(t)}{r_a(t)}  - \frac{5}{2\sqrt{2}}\frac{M(t)^{\frac{3}{2}}}{r_a(t)^{\frac{3}{2}}} +{\cal O}\left(\frac{M(t)^2}{r_a(t)^2}\right)\right]\,,\label{rc}
\end{eqnarray}
in a time domain where $M(t)\ll r_a(t)$ drawing the conclusion that when the time is increasing in  expanding universes the cosmological horizon tends to the asymptotic one while the black hole horizon is shrinking to the sphere of radius $2 M(t)$ collapsing then to zero (as in Fig. 2).  Note that this property holds even in the case when the asymptotic space-time is the de Sitter expanding universe shown in the right panel of Fig. 2.  

\begin{table}
	\begin{tabular}{ccc}
		\hline
		&&\\
		&{ McVittie dyn.  particles}&{Schwarzschild-type dyn.  particles}\\	
		&&\\
		asympt. space-time&any FLRW  \cite{MV4}&spatially flat  FLRW  ${\frak M}(a)$\\
		&&\\
		static limit&distorted as in Eq. (\ref{asymMv})&natural as in Eq. (\ref{s}) with $f=f_S$\\
		&&\\
		space sections& curved& flat\\
		&&\\
		particle mass&static,  $m$&dynamical, $M(t)=M_0 a(t)^{-3}$\\
		&&\\
		black hole horizon& $r_b(t) \to 2m$ (as in Fig. 1)& $r_b(t)\to 2M(t)\to0$  (as in Fig. 2)\\
		&&\\
		cosm. horizon&$r_c(t)\to r_a(t)$&$r_c(t)\to r_a(t)$\\
		&&\\
		fluid density& unchanged&  singular in origin \\
		&&\\
		fluid pressure& singular at $r=2m$& unchanged\\
		&&\\
		\hline
	\end{tabular}
	\caption{Comparation between McVittie  and Schwarzschild-type dynamical particles with expanding asymptotic FLRW spacetimes.  }
\end{table}

Comparing  the McVittie and Schwarzschild-type dynamical particles as in Tab. 1 we may conclude that these are systems with different dynamics despite of some general common features as, for example, the existence of the cosmological and black hole horizons.   Recently the McVittie dynamical particles have been generalized to manifolds allowing asymptotic FLRW space-times with curved space sections \cite{MV4}. Therefore we may ask how the Schwarzschild-type ones can be generalized in a non-trivial manner as new solutions of Einstein's equations with perfect fluid  with singular densities  but without abandon the spatially flat asymptotic FLRW space-times.

\section{New solutions of Einstein's equations with perfect fluid}

 In what follows we try to find new solutions of Einstein's equations in space-times with perfect fluid assuming that these must have: line elements  (\ref{s1}) with flat space sections (i. e. $g_{rr}=1$),  the asymptotic form (\ref{s2}) and  static limit as in Eq.  (\ref{ss}).

As in Ref. \cite{Cot} we consider a class of metrics  of the form (\ref{fam}) but with  more general functions 
\begin{equation}
	h(t,r)=N(t) r+\sqrt{\frac{2 M(t)}{r}+\Omega(t)^2 r^2}\,,
\end{equation}
depending on the dynamical masses, $M(t)$, which are arbitrary functions of time representing the principal dynamical parameters. The functions $N(t)$ and $\Omega(t)$ results from Eqs. (\ref{Ein})  as
\begin{equation}
	N(t)=-\frac{1}{3M(t)}\frac{dM(t)}{dt}=-\frac{1}{3}\frac{\dot{M}(t)}{M(t)}\,, ~~~~~\Omega(t)=\kappa M(t)\,,
\end{equation}  
where $\kappa\in \mathbb{R}$ is an integration constant playing the role of free parameter. We obtain thus the definitive form of the $h$-function, 
\begin{equation}\label{hdef}
	h_{\kappa}(t,r)=-\frac{1}{3}\frac{\dot{M}(t)}{M(t)} r+\sqrt{\frac{2 M(t)}{r}+\kappa^2 M(t)^2 r^2}\,,
\end{equation}
defining the line elements (\ref{fam}) of the new apace-times we denote from now by ${\frak M}(M,\kappa)$ referring them as $\kappa$-models. The Einstein tensors of these models have in their proper physical frames the isotropic form 
\begin{eqnarray}
G^0_0(M,\kappa)&=&8\pi \rho_{\kappa}=3\kappa^2 M(t)^2 +\frac{1}{3} \frac{\dot M(t)^2}{M(t)^2}-2\dot{M}(t)\frac{1+{M(t)}\kappa^2r^{3} }{\sqrt{M(t)(M(t)\kappa^2r^3+2) r^3}}\,,\\
G(M,\kappa)&=&-8\pi p_{\kappa}=3\kappa^2 M(t)^2+ \frac{\dot M(t)^2}{M(t)^2}-\frac{2}{3} \frac{\ddot M(t)}{M(t)}	\,,
\end{eqnarray}
satisfying the condition (\ref{cond}). Thus we may say that the $\kappa$-models defined here are exact solutions of Einstein's equations with perfect fluid of density $\rho_{\kappa}$ and pressure $p_{\kappa}$. 

For any such model the form of the function (\ref{hdef})  guarantees an asymptotic FLRW space-time ${\frak M}(a)$ whose Hubble function satisfies
\begin{equation}\label{at}
	\frac{\dot a(t)}{a(t)}=\lim_{r\to\infty}\frac{ h(t,r)}{r}=\kappa M(t) -\frac{1}{3}\frac{\dot{M}(t)}{M(t)}\,.
\end{equation}
The Hubble function must be a monotonously time dependent function without zeros in the physical time domain since these might produce singularities of the function $r_a(t)$ giving the radius of the asymptotic horizon.  We may prevent  these zeros to appear in two cases, either for expanding geometries when we must take
\begin{equation}\label{expand}
\frac{\dot a(t)}{a(t)}>0 ~~\Rightarrow ~~ 	\frac{\dot{M}(t)}{M(t)}<0\,,\quad \kappa>0\,,
\end{equation} 
or for collapsing ones for which we have to chose
\begin{equation}
	\frac{\dot a(t)}{a(t)}<0 ~~\Rightarrow ~~ 	\frac{\dot{M}(t)}{M(t)}>0\,,\quad \kappa<0\,.
\end{equation} 
In what follows we restrict ourselves to the expanding space-times which are of interest in cosmology. Therefore, we have to consider mass functions decreasing monotonously in time and positive parameters $\kappa\ge 0$. 

Integrating then Eq. (\ref{at})  with the initial condition $a(t_0) =1$ we obtain the scale factor  
\begin{equation}\label{at1}
a(t)=\left(  \frac{M_0}{M(t)}\right)^{\frac{1}{3}} \exp\left( \kappa\int_{t_0}^t M(t')dt'\right)
\end{equation}
of the asymptotic FLRW space-time  where $M_0=M(t_0)$ such that we preserve the initial condition  applied previously to the function (\ref{lineBH}). 

On the other hand, we observe that we may reconstruct these space-times starting with the parameters $a(t),\,M_0$ and $\kappa$ recovering the mass function as
\begin{equation}\label{mat} 
	M(t)=\frac{M_0}{a(t)^3}\left[ 1-3\kappa M_0 \int_{t_0}^t \frac{dt'}{a(t')^3}\right]^{-1}\,,
\end{equation} 
and bringing the $h$-function in the form
\begin{eqnarray}\label{hdef1}
	h_{\kappa}(t,r)=\left(\frac{\dot a(t)}{a(t)}-\kappa M(t)\right) r+\sqrt{\frac{2 M(t)}{r}+\kappa^2 M(t)^2 r^2}\,.
\end{eqnarray}
Under such circumstances the $\kappa$-models   may be seen as perturbations of the perfect fluid of their asymptotic space-times. Writing the Einstein tensors in co-moving physical frames as
\begin{eqnarray}
	G^0_0(M,\kappa)&\equiv&G^0_0(a,M_0,\kappa)=G^0_0(a)+\delta_{\kappa}\,,\\
	G(M,\kappa)&\equiv&G(a,M_0,\kappa)=G(a)\,,
\end{eqnarray}
we understand that these dynamical particles modify only the fluid density preserving its pressure. These perturbations are given by the term 
\begin{equation}
	\delta_{\kappa}(t,r)=2 \dot M(t)\left[  \kappa  -\frac{1+{M(t)}\kappa^2r^{3} }{\sqrt{M(t)(M(t)\kappa^2r^3+2) r^3}}   \right] \,,	
\end{equation}
which is singular in $r=0$ and satisfies the expected asymptotic condition $\lim_{r\to\infty} \delta(t,r) =0$.

When we use the parameters $a(t)$, $M0$ and $\kappa$ we denote  the space-times by ${\frak M}(a,M_0,\kappa)$ bearing in mind that this is isomorphic with ${\frak M}(M,\kappa)$ through Eqs. (\ref{at1}) and (\ref{mat}) which relate their parameters. For $\kappa=0$ we recover our point dynamical particles \cite{Cot},  
\begin{equation}
{\frak M}(M,0) \sim {\frak M}(a,M_0,0)={\frak M}(a,M_0)	\,,
\end{equation}
with  $h$-functions of the form (\ref{lineBH}). However, when we construct the space-times ${\frak M}(a,M_0,\kappa)$ starting with the asymptotic scale factor $a(t)$   we must prevent the possible poles in Eq. (\ref{mat}) imposing the restriction  $\kappa\in[0,k_{lim}]$ where $k_{lim}$ must satisfy
\begin{equation}\label{coco}
 3\kappa_{lim} M_0 \int_{t_0}^\infty \frac{dt'}{a(t')^3}	=1\,,
\end{equation}  
because the function $a(t)$ is positively defined.

Regardless the parameters we use,  the $\kappa$-models describe dynamical particles which behave as black holes inside their physical domains bordered by the spheres of black hole and cosmological horizons. The radii of these horizons have to be derived by solving the equation $f_{\kappa}(t,r) =1$ for expanding universes or $f_{\kappa}(t,r) =-1$ for collapsing ones. In both these cases we have to solve cubic equations according to the method presented in the Appendix. The real and positive solutions $0<r_b(t) < r_c(t)$ represent the radii of the black hole and respectively cosmological horizons which  arise at the critical time $t_{cr}$ when $\Delta(t_{cr})=0 \Rightarrow    r_b(t_{cr}) =r_c(t_{cr})$ evolving then in time as C-curves (as in Figs. 3 and 5). In general, for $\kappa\not=0$ these functions are complicated and less intuitive. Moreover, these cannot be expanded as in Eqs. (\ref{rb}) and (\ref{rc}) because of the parameter $\kappa$ which can take arbitrary values. Therefore, for understanding the principal features of the $\kappa$-models we must resort to numerical and graphical analyses. 

\section{Simple examples}

 { \begin{figure}
		\centering
		\includegraphics[scale=0.34]{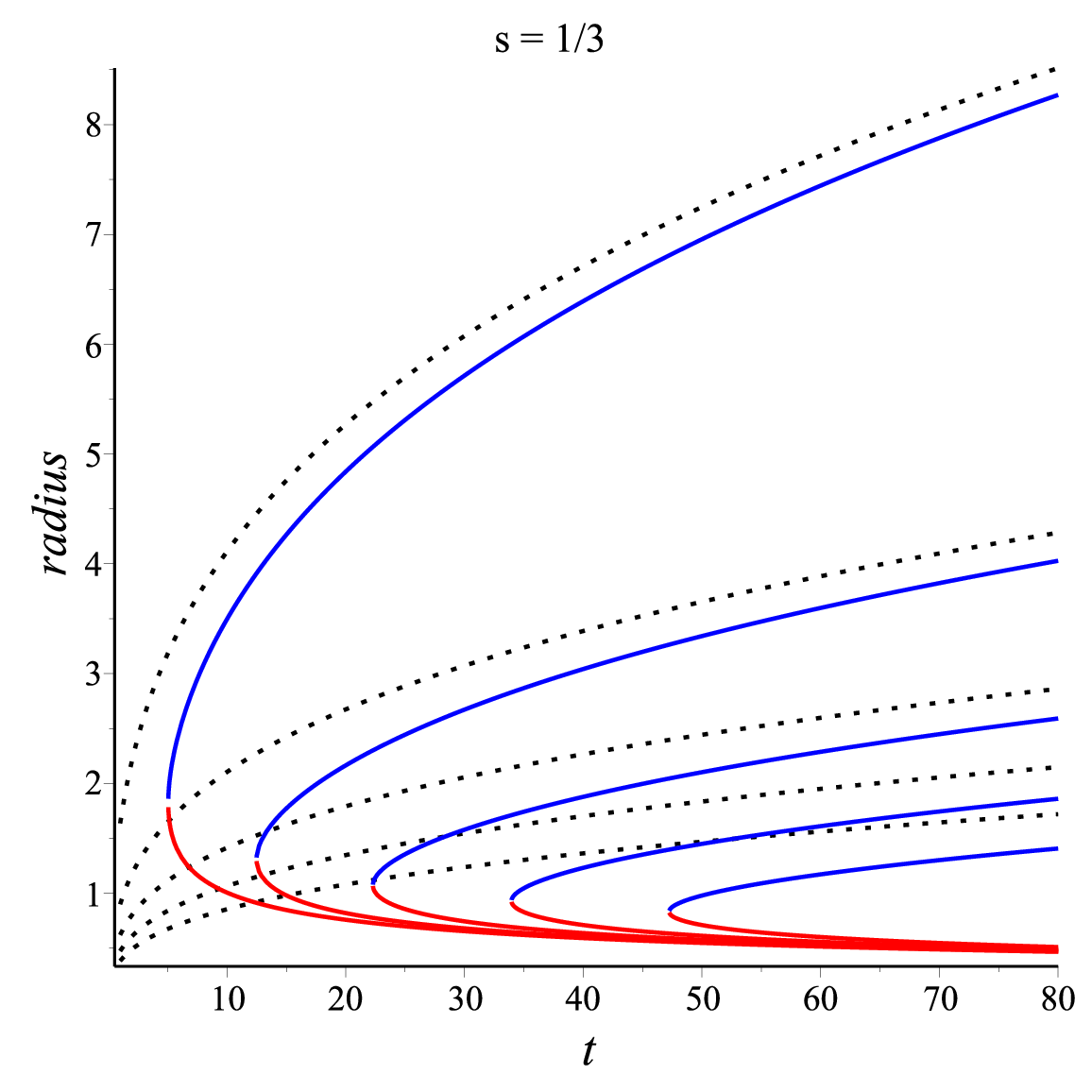}
		\includegraphics[scale=0.34]{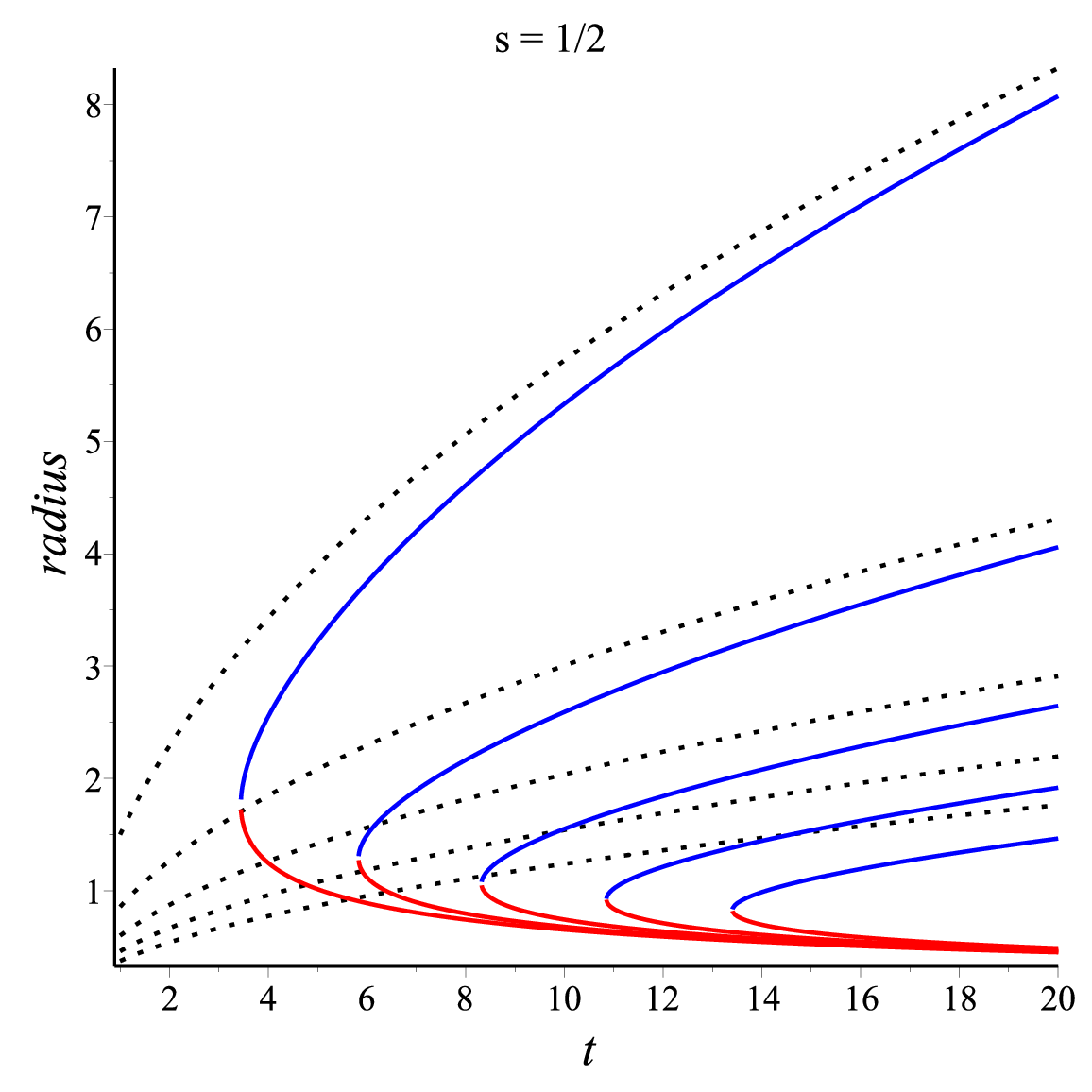}	
		\includegraphics[scale=0.34]{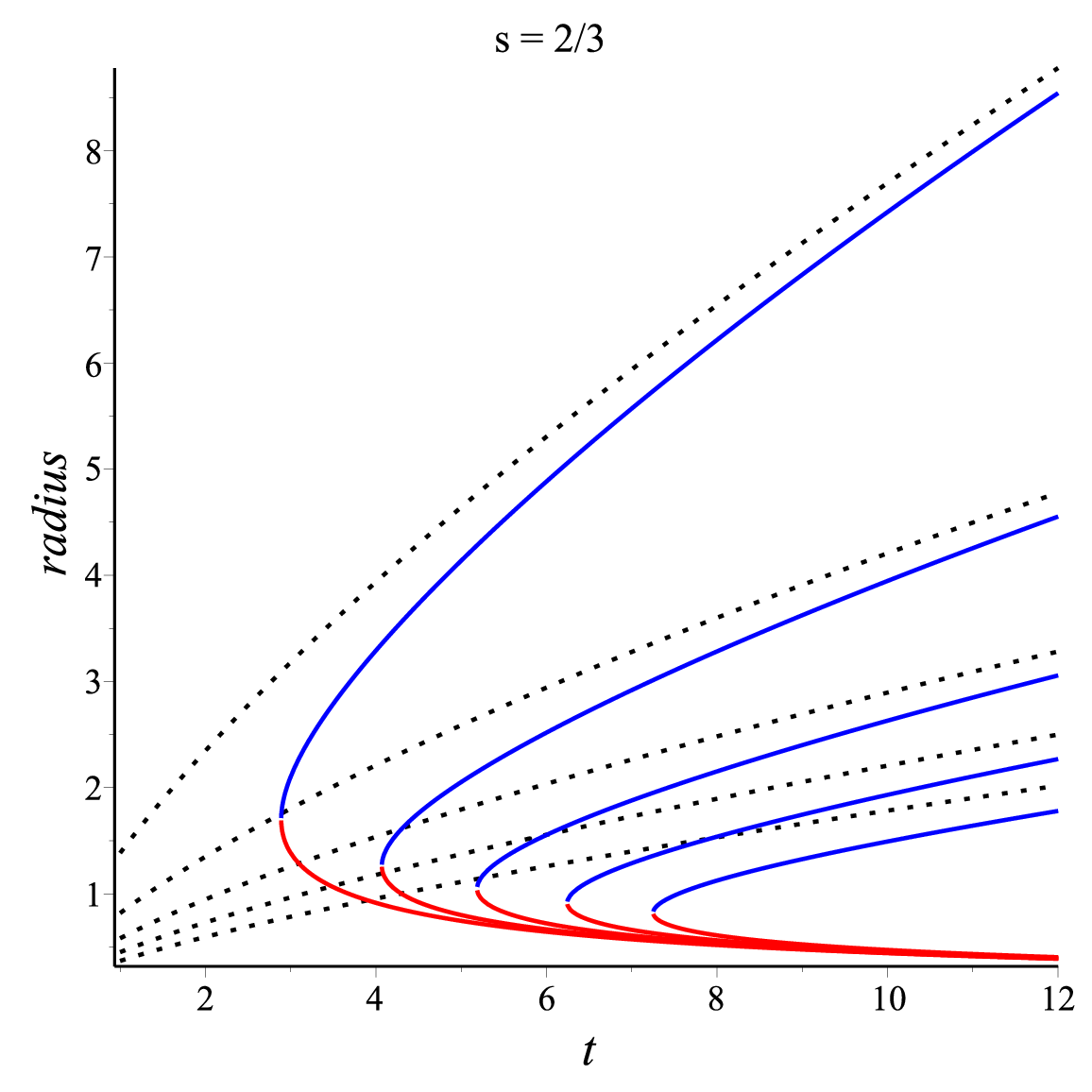}	
		\includegraphics[scale=0.34]{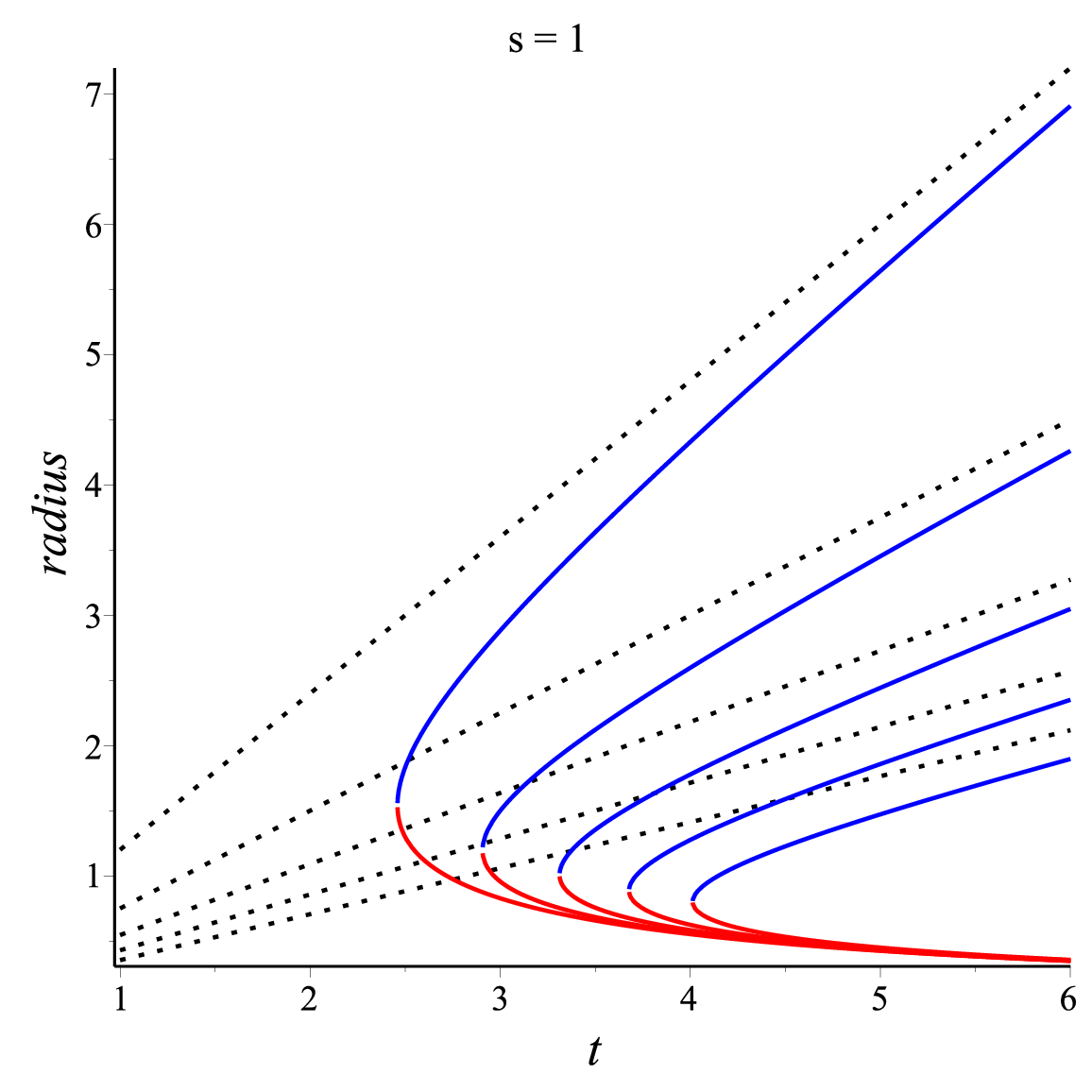}		
		\caption{Evolution of the radii of the horizon spheres of  models $A(s,\kappa)$.  The radii of the asymptotic (dotted), cosmological (upper solid) and black hole (lower solid) horizons are plotted for different values  of $s$ and  $\kappa=0.5, 1, 1.5,2,2.5$  corresponding to  the C-curves in descending order. }
\end{figure}}  

Let us focus now on the simplest examples that could point out the role of the parameter $\kappa$ of our new models proposed here without exceeding the performance of usual PC-s in numerical calculations. We restrict ourselves to the $\kappa$-models of expanding space-times which  comply with the conditions (\ref{expand}).

{\em Example A} We consider first the models  of dynamical particles we denote by $A(s,\kappa)$ defined in space-times ${\frak M}(M,\kappa)$ with
\begin{equation}
	M(t)=M0\left( \frac{t0}{t} \right)^s\,,\quad \kappa\in \mathbb{R}^+\,,
\end{equation}
assuming that $s\ge 0$. In the trivial case of $s=0$, the mass is static,  $M(t)=M_0$, and the metric becomes a Schwarzschild-de Sitter one with the line elements in physical frames of the form (\ref{frSdS}) but with $m=M_0$ and the Hubble de Sitter constant $\omega_{SdS}=\kappa M_0$. The genuine dynamical $A$-models have to be obtained  for $s>0$ when the asymptotic behaviour is of FLRW type. The scale factors of the asymptotic space-times  derived from Eq. (\ref{at1}) give the radii of the asymptotic horizons (\ref{hor}) as
\begin{eqnarray}
a(t)=\left(\frac{t}{t_0}\right)^{\frac{s}{3}} \exp\left[\kappa M_0t_0^{s}\int_{t_0}^t\tau^{-s} d\tau\right]~~~ \Rightarrow~~~  r_a(t)=3\frac{t}{t_0}\left[ s+3\kappa M_0\left(\frac{t}{t_0}\right)^{1-s}\right]^{-1}\,.
\end{eqnarray}
Deriving then the horizon radii (\ref{rbc}) we may plot in Fig. 3 the evolution of the  horizon radii of the $\kappa$-models for various values of parameters $s$ and $\kappa$, taking $M_0=1$ and $t_0=1$.  These plots show the role of the parameter $\kappa$ which determines different asymptotic behaviours of models having the same mass function. Moreover, this parameter is responsible for the profile of the fluid density which carries the singularity in $r=0$ that remains outside the physical domain. Nevertheless, the density in the physical domain feels its effect  increasing near the black hole horizon as in Fig. 4. However,  for larger values of $\kappa$ the density tends to homogeneity (as in the right panel of Fig.  4) which means that this parameter is able to suppress the effect of the singularity when this is increasing enough. 

{ \begin{figure}
		\centering
		\includegraphics[scale=0.34]{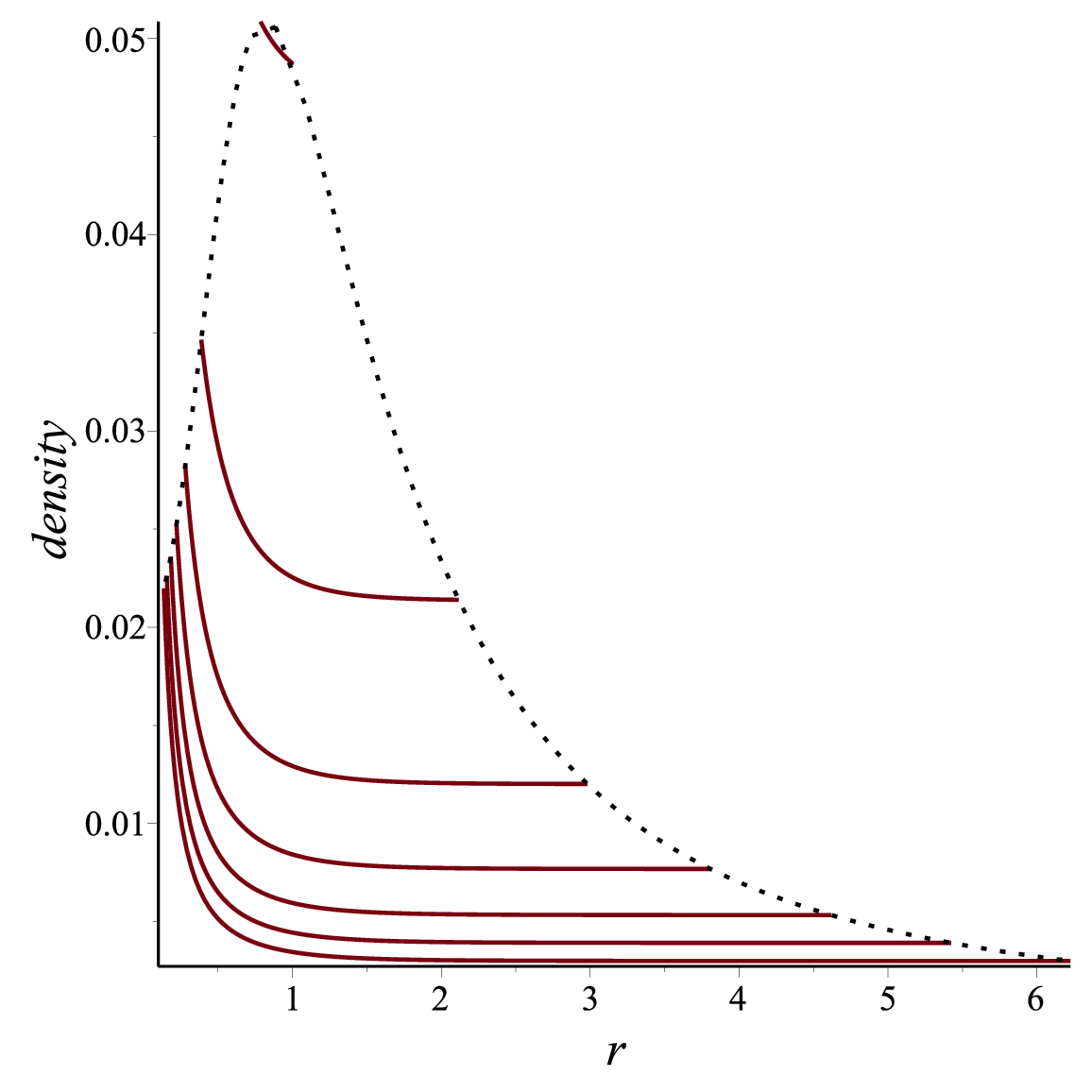}
		\includegraphics[scale=0.34]{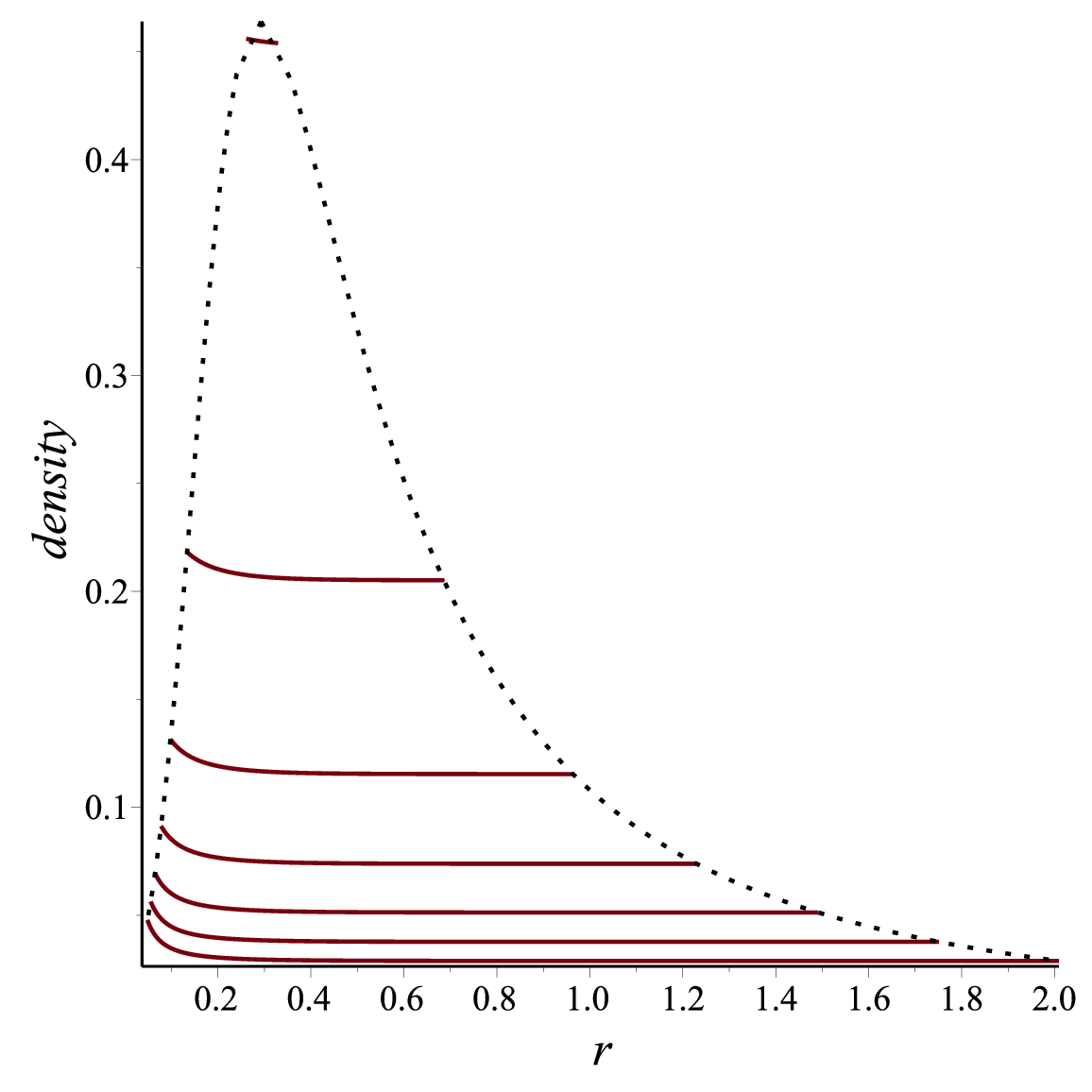}	
		\caption{Density profiles of models $A(s,\kappa)$ with $s=1$ and  $\kappa=2$ (left panel) and $\kappa=20$ (right panel). The functions $\rho_{\kappa}(t,r)$ are plotted  inside the physical domain (between the black hole and cosmological horizons) at different instants,  $t=1.001, 1, 1.5, 2, 2.5, 3$  (in units of $t_{cr}$) corresponding to the plotted profiles in descending order.  }
\end{figure}}

{\it Example B}  The $\kappa$-models   can be constructed using te parameters $a(t)$, $M_0$ and $\kappa$ for writing down the metric tensors of the space-times ${\frak M}(a,M_0,\kappa)$. In this manner we may study families of models  having the same asymptotic behaviour determined by the function $a(t)$. Here we consider the simple case of  models denoted by $B(p,\kappa)$ having the scale functions
\begin{equation}
	a(t)=\left( \frac{t}{t_0}\right)^p\,, \quad p>0\,,
\end{equation}
which satisfy our initial condition $a(t_0)=1$. In these models the mass functions may be derived as in Eq. (\ref{mat}) but respecting the restriction $0\le\kappa\le\kappa_{lim}$ where 
\begin{equation}
\kappa_{lim}=\frac{3p-1}{3M_0t_0}\,,\quad  p>\frac{1}{3}\,,	
\end{equation}
as it results from  Eq. (\ref{coco}). This condition guarantees that the mass functions of these models
\begin{equation}
	M(t)=\frac{M_0t_0}{t}\left[ \left(\frac{t}{t_0}\right)^{3p-1}\left(1-\frac{3\kappa M_0t_0}{3p-1}\right)+\frac{3\kappa M_0t_0}{3p-1}\right]^{-1}\,,
\end{equation}
do not have singularities  but restricts severely the values of the parameter $\kappa$. For this reason the models of families with the same asymptotic horizon are very close to each other having horizons whose radii form very similar C-curves as in Fig. 5. Note that we can set $p=\frac{1}{3}$ only when $\kappa=0$ and  the models $A(3p,0)\equiv B(p,0)$ with any $p>0$  become  the dynamical particles of Ref. \cite{Cot}.  

In other respects, we observe that the parameter $\kappa$ gives some flexibility to our approach such that we may find non-trivial equivalences between different types of models as in the case of the models $A(s=1,\kappa) \equiv B(p=\frac{1}{3}+\kappa M_0 t_0, \kappa)$ which are equivalent for any $\kappa\ge 0$ if we  set the same values of the constants $M_0$ and $t_0$. 

 { \begin{figure}
		\centering
		\includegraphics[scale=0.34]{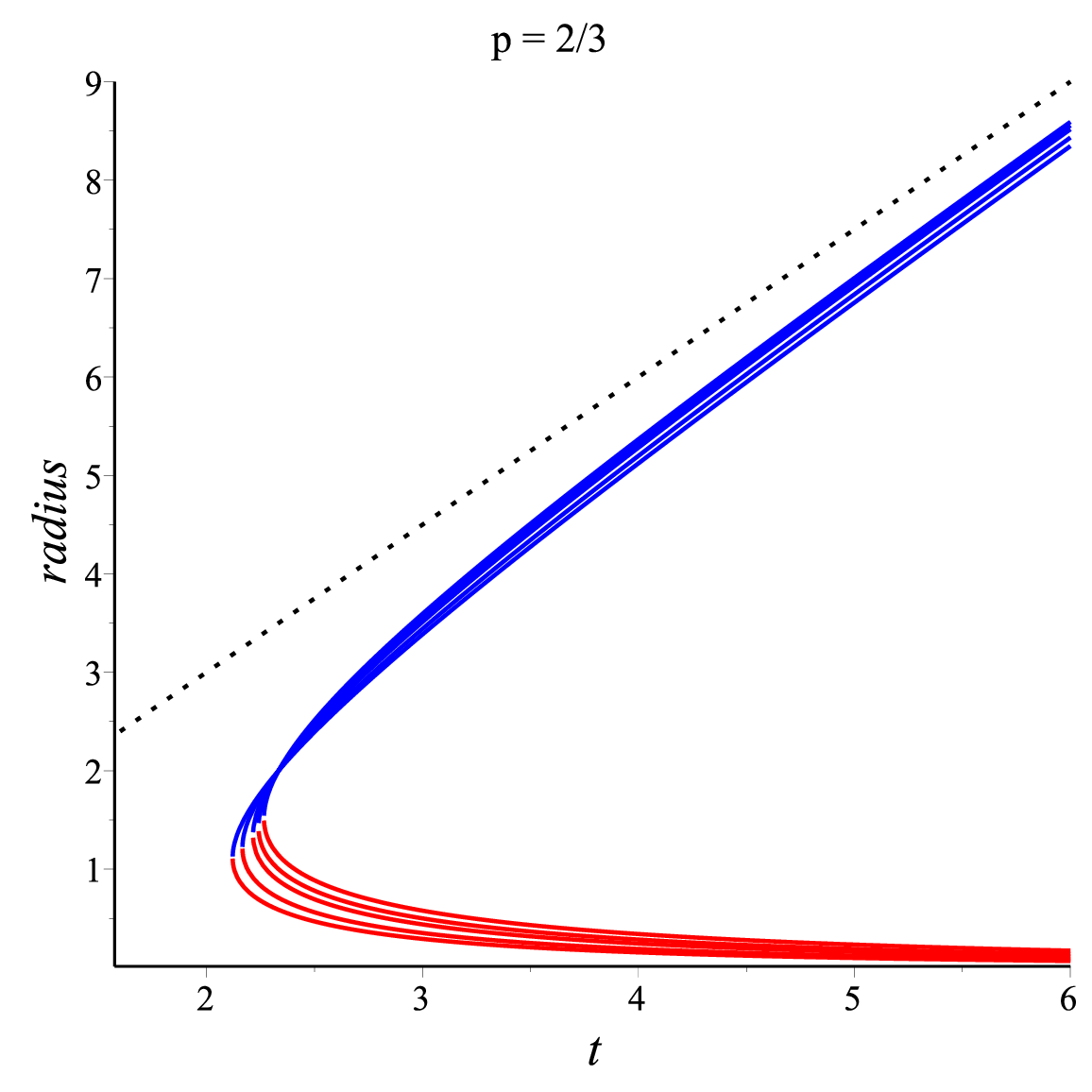}
		\includegraphics[scale=0.34]{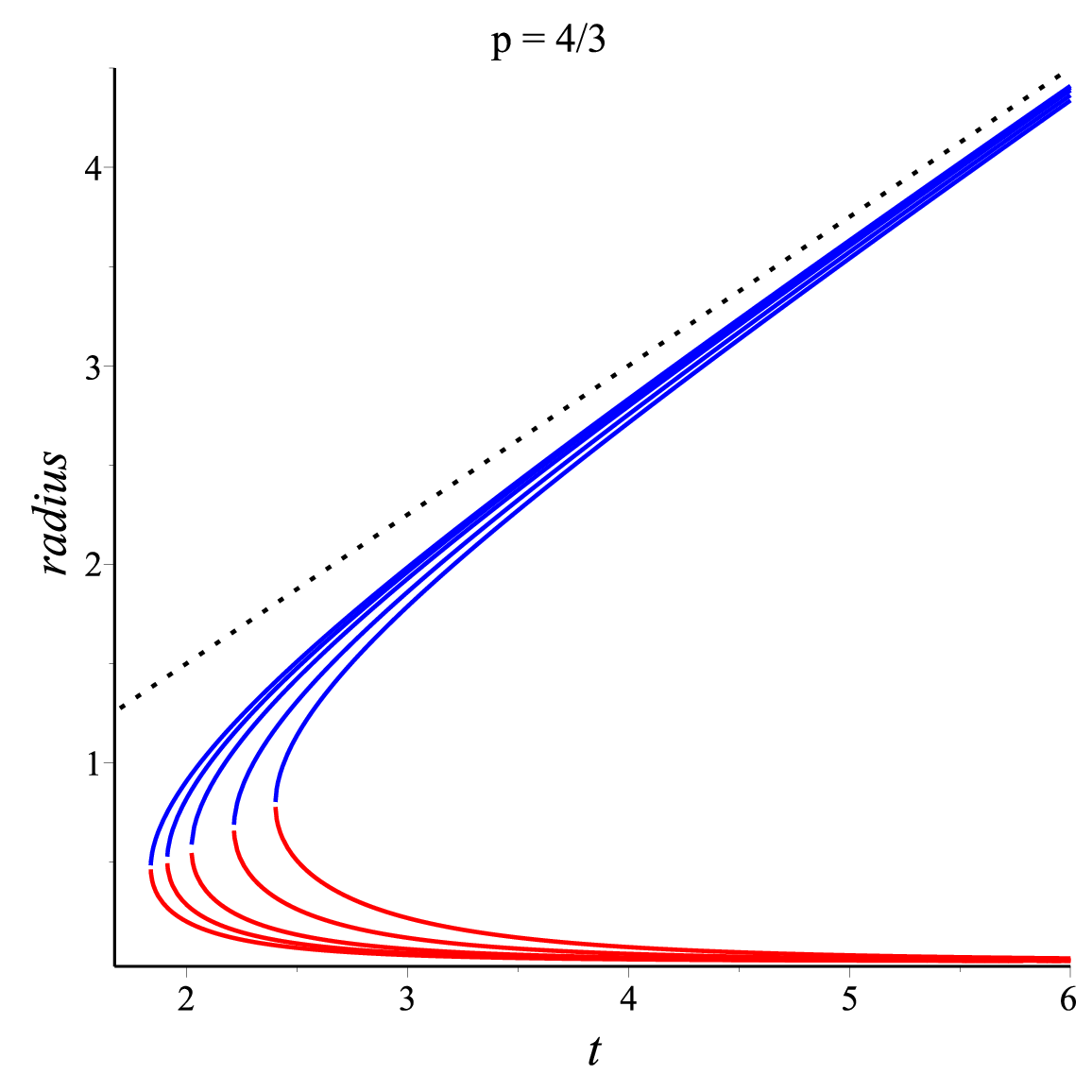}	
		\caption{Two families of models $B(p,\kappa)$ with asymptotic FLRW space-times with $a(t)=t^p$ and $M_0=t_0=1$. Five C-curves are plotted corresponding (from left to right)  to five equidistant  values of $\kappa$ in the interval $[0,\kappa_{lim}]$.  }
\end{figure}}

\section{Concluding remarks}

We proposed here a new family of space-times depending on a free parameter showing that these represent  dynamical particles produced by central singularities of the fluid density but without affecting the pressure of the  perfect fluid of the asymptotic FLRW space-times.  Each such  dynamical particle becomes a black hole at a critical instant $t_{cr}$ when a pair of black hole and cosmological horizons appear on the same sphere.   For $t>t_{cr}$ these horizons evolve creating the physical space between their spheres where a remote observer may perform measurements. 

All these models depend on the free parameter $\kappa$ resulted as an integration constant when we integrated the Einstein equations. The mass function and  this parameter are enough for defining the model determining the time evolution of the entire manifold including the evolution of the cosmological and black hole horizons  as well as the scale factor of the asymptotic FLRW space-time. An advantage of our  $\kappa$-models is that these can be constructed using an alternative set of  parameters formed by the scale factor of the asymptotic FLRW space-time, initial mass and  $\kappa$. Regardless the parameters we use $\kappa$ remains the typical  parameter of our new models.  When this parameter vanishes we recover the dynamical particles of Ref. \cite{Cot} which appear now as a particular cases of the present approach. 

As in Ref. \cite{Cot} we argued that the models with $\kappa=0$ represent a new type of dynamical particles we may conclude that the $\kappa$-models with $\kappa\not= 0$ we proposed and studied here are new dynamical particles that could populate new interesting models of dynamical universes.

\appendix

\section{Solving cubic equations}

\setcounter{equation}{0} \renewcommand{\theequation}
{A.\arabic{equation}}

For solving the cubic equation 
\begin{equation}
ar^3+br^2+cr+d=0\,, \label{e1}	
\end{equation}
 we substitute first  $r=x-\frac{b}{3a}$ obtaining the modified depressed equation $x^3-p x+q=0$ with the coefficients
\begin{eqnarray}
	p=\frac{b^2-3ac}{3a^2}\,,\quad q=\frac{27a^2d-9abc+2b^3}{27a^3}\,,
\end{eqnarray}
that can be solved using the following form of  Cardano's formulae 
\begin{eqnarray}
	x_1&=&\frac{1}{2}\left(A+\frac{4}{3}\frac{p}{A}\right)\,,\\
	x_2&=&\frac{1}{2}\left(Ae^{i\frac{2\pi}{3}}+\frac{4}{3}\frac{p}{A}e^{-i\frac{2\pi}{3}}\right)\,,\\
	x_3&=&\frac{1}{2}\left(Ae^{i\frac{4\pi}{3}}+\frac{4}{3}\frac{p}{A}e^{-i\frac{4\pi}{3}}\right)\,,	
\end{eqnarray}
where $	A=\left(\frac{2}{3}\right)^{\frac{2}{3}}\left[ i\sqrt{3}\sqrt{4p^3-27 q^2}-9 q \right]^{\frac{1}{3}}$. The cubic equations allow real solutions  only when their discriminants are positive, $\Delta=4 p^3-27 q^2 >0$. 

The horizons of the McVittie dynamical particles result directly from the depressed equation
\begin{equation}
g_{00}(t,r)=0~~\Rightarrow~~x(t)^3-x(t)+\frac{2m}{r_a(t)}=0\,, \quad x(t)=\frac{r(t)}{r_a(t)}\,,	
\end{equation} 
which has real solutions for $t>t_{cr}$ where $t_{cr}$ solves the equation $\Delta(t_{cr})=0$ obeying $r_a(t_{cr})=3\sqrt{3}\,m$. The black hole and cosmological horizons are given by $r_b(t)=r_a(t)\, x_3(t)$ and $r_c(t)=r_a(t)\, x_1(t)$.

For our $\kappa$-models  the cubic equations $h_{\kappa}(t,r)=\pm1$ can be put in the canonical form (\ref{e1}) with the coefficients 
\begin{equation}
a(t)=\nu(t)^2-\lambda(t)^2\,, ~~~~b(t)=2\nu(t)\,,~~~~c=\pm 1\,,~~~~d(t)=-\mu(t)\,,
\end{equation}
where we denote
\begin{equation}
\lambda(t)=\kappa M(t)\,,\quad\mu(t)=2M(t)\,,\quad \nu(t)=\frac{1}{3}\frac{\dot M(t)}{M(t)}\,,	
\end{equation}
taking into account that the function $h_{\kappa}(t,r)$ is defined by Eq. (\ref{hdef}). Finally the solutions we are looking for can be identified as
\begin{equation}\label{rbc}
r_b(t)=x_1(t)-\frac{b(t)}{3a(t)}\,,\quad 	r_c(t)=x_3(t)-\frac{b(t)}{3a(t)}\,,\quad \forall t\ge t_{cr}\,,
\end{equation}
while $x_2(t)<0$ is the nonphysical solution. Obviously, for $\kappa=0$ we recover the results of Ref. \cite{Cot}.

\end{document}